\documentclass{JHEP}

\newcommand{\be}{\begin{equation}}
\newcommand{\ee}{\end{equation}}

\newcommand{\bea}{\begin{eqnarray}}
\newcommand{\eea}{\end{eqnarray}}
\newcommand{\bean}{\begin{eqnarray*}}
\newcommand{\eean}{\end{eqnarray*}}

\def\beq{\begin{equation}}

\def\eeq{\end{equation}}

\def\Tr{\mathop{\rm Tr}}

\def\eref#1{(\ref{#1})}

\def\vol{\mathop{\rm Vol}}

\preprint{UPR-1022-T\\ {\tt hep-th/yymmddd}}
\title{Seiberg Duality in Matrix Models II}

\author{Bo Feng$^1$ and Yang-Hui He$^2$\\
$^1$Institute for Advanced Study \\
Einstein Drive, \\
Princeton, New Jersey, 08540 \\
~\\
$^2$Department of Physics,\\
The University of Pennsylvania,\\
Philadelphia, PA 19104-6396
}
\abstract{In this paper we continue the investigation, within the context of
the Dijkgraaf-Vafa Programme, of Seiberg duality
in matrix models as initiated in hep-th/0211202, 
by allowing degenerate
mass deformations. In this case, there are some massless fields which
remain and the theory has a moduli space. 
With this illustrative example, we propose a general
methodology for performing the relevant matrix model integrations and
addressing the corresponding field theories which have non-trivial IR
behaviour, and which may or may not have tree-level superpotentials.
}
\begin{document}
\section{Introduction}
With rapid development, the Dijkgraaf-Vafa Programme
\cite{Dijkgraaf:2002fc,Dijkgraaf:2002vw,Dijkgraaf:2002dh}
of establishing a
correspondence between a wide class of four dimensional ${\cal N}=1$
gauge theories and certain bosonic matrix models has withstood
extensive tests (q.~v.~ \cite{Dorey:2002jc} to \cite{Feng}). The
original proposal was that an ${\cal N}=1$ $U(N_c)$ gauge theory with
a tree level superpotential $W_{tree}(\Phi_i, g_a)$ in adjoint fields
$\Phi_i$ and couplings $g_a$ has a complete effective superpotential
\begin{eqnarray*}
W_{eff}(S,\Lambda,g_a) &=& W_{\mbox{Veneziano-Yankielowicz}} +
	W_{\mbox{Perturbative}} \\
&=&N_c S (1 - \log(\frac{S}{\Lambda^3})) + N_c \frac{\partial {\cal F}_0(S,
	g_a)}{\partial S}
\end{eqnarray*}
where
$S := -\frac{1}{32 \pi^2} \Tr W_a W^a$ is the glueball superfield,
$\Lambda$, the cutoff scale, and ${\cal F}_0$ is the genus-zero (i.e.,
planar) partition function (at large rank $M$) 
of the matrix model whose potential is
formally the tree-level superpotential:
\begin{eqnarray*}
{\cal F}_0 := {\cal F}_{\chi = 2} &=& -\frac{S^2}{M^2} \log Z \\
&=& -\frac{S^2}{M^2} \log \int [D\Phi_i] \exp\left( -\frac{M}{S}
W_{tree}(\Phi_i,g_a) \right).
\end{eqnarray*}

The addition of flavour to the above story has also been performed
\cite{Argurio:2002xv,McGreevy:2002yg,Suzuki:2002gp,Bena:2002kw,Demasure:2002sc}.
Let us adhere to the conventions of \cite{Argurio:2002xv}. Now an
$U(N_c)$ ${\cal N}=1$ theory with adjoint $\Phi$ and $N_f$
fundamentals $Q_f$ and $\widetilde{Q}^f$ with tree-level superpotential
$W_{tree}(\Phi_i, Q_f, \widetilde{Q}^f, g_a)$ has, according to the
correspondence, the effective superpotential
\begin{eqnarray}
\label{flavour}
W_{eff}(S,\Lambda,g_a) &=& W_{\mbox{Veneziano-Yankielowicz}} +
	W_2 + W_1 \nonumber \\
&:=& N_c S (1 - \log(\frac{S}{\Lambda^3})) + N_c \frac{\partial {\cal
	F}_{\chi=2}(S,g_a)}{\partial S} + {\cal F}_{\chi=1}(S,g_a)
\end{eqnarray}
where
\[
-\frac{M^2}{S^2} {\cal F}_{\chi = 2}(S,g_a) - \frac{M}{S} {\cal
 F}_{\chi = 1}(S,g_a) = \log \int [D\Phi DQ_f D\tilde{Q}^f] \exp
 \left( -\frac{1}{g_s} W_{tree}(\Phi_i, Q_f, \widetilde{Q}^f, g_a) \right).
\]
In other words, ${\cal F}_{\chi = 2}(S,g_a)$ is the genus zero 
planar contribution and ${\cal F}_{\chi = 1}(S,g_a)$, the boundary 
contribution from flavours. 
We remark that in these above computations the matrix model is of rank
$M$ (which is to be taken to
infinity), a parametre unrelated to $N_c$ and $N_f$.

With these pieces of information, together with the already existent
literature on the full non-perturbative pure $U(N_c)$ SUSY gauge
theory, viz., the Affleck-Dine-Seiberg superpotential \cite{ADS}, an
immediate check presents itself to us, namely Seiberg Duality
\cite{Seiberg:1994pq}. This was done by the first author in
\cite{Feng}.

In particular the check was performed thus. The archetypal example of
a Seiberg dual pair wherein both the electric and magnetic sides have
tree level superpotentials involve mass deformations of the following
type
\[
\begin{array}{|c|c|c|}
\hline
\mbox{Electric} & \stackrel{\mbox{Seiberg}}{\leftrightarrow} & \mbox{Magnetic} \\
\hline
W_{ele} = \sum\limits_{i=1}^{N_f} Q_j m_j \widetilde{Q}^j & &
W_{mag} = \sum\limits_{i=1}^{N_f} \frac{1}{\mu} X_i^j q_j \tilde{q}^j + \Tr
	(X m) \\ \hline
\multicolumn{3}{|c|}{W_{eff} = N_c(\hat{\Lambda}^{3N_c})^{\frac{1}{N_c}}
= N_c(\Lambda^{3N_c-N_f}\det(m))^{\frac{1}{N_c}} }
\\ \hline
\end{array}
\]
where $m$ are the non-degenerated mass matrix 
of the fundamental squarks $Q_j, \widetilde{Q}^j$ on the electric
side (the matrix $m$ is thus diagonal with entries $m_j$), 
$\mu$, a dynamical scale, $\Lambda$, the UV cutoff scale,
$\hat{\Lambda}$, the IR cutoff scale, $q_j, \tilde{q}^j$, the dual
quarks and $X$, the dual meson. In \cite{Feng} the matrix model
computations were carried out for $W_{ele}$ and $W_{mag}$ individually
according to (\ref{flavour}), and $W_{eff}$ was retrieved for both,
whereby beautifully supporting the validity of the Dijkgraaf-Vafa
Correspondence once more.

The story however, is not complete. To fully understand Seiberg
duality one needs to consider cases without mass
deformations, and thus indeed
flat directions in the moduli space. This is to
say that whereas \cite{Feng} addressed the case where the mass matrix
$m$ was maximal rank, we need to explore the more subtle case when $m$
has zero eigenvalues. This purpose of this writing is to supplant
the analysis of \cite{Feng} by showing that in this case of flat
directions the Dijkgraaf-Vafa Programme continues to hold. It is
important to study this example because it enpowers us with techniques
as to what to do when the field theory has a non-trivial IR moduli
space; moreover, theories which have no tree-level superpotentials
which seem {\it per definitio} to elude the Programme can be
treated by certain addition of appropriate constraints.

The organization of this paper is follows. We commence with
a short review of the field theory results in Seiberg Duality in
Section 2. Then, in Section 3,
we directly integrate the corresponding electric and magnetic 
matrix models and show that they reproduce the known field theory
results whereby supporting the Dijkgraaf-Vafa Programme
for Seiberg duality in this more general and illustrative case.
Finally, in Section 4, we give a discussion on some 
interesting problems and prospects.
%
\section{A Review of the Field Theory}
The phase structure of $SU(N_c)$ SUSY gauge theory with $N_f$ flavors 
$Q_i, \widetilde{Q}_i$ and no superpotential has been analyzed
in \cite{Intriligator:1994sm} (q.~v.~ \cite{Argyres} for a more
pedagogic review).
In general the theory will have a moduli space described by gauge invariant 
operators, namely the meson field
$M_i^j= Q_i\widetilde{Q}_j$ and the baryon fields
$B$ and $\widetilde{B}$.
The baryons exist only when $N_f\geq N_c$ because they are constructed as
being totally antisymmetric in the color index. 

In the case at hand,
$M_i^j, B, \widetilde{B}$
are not independent and satisfy some constraints, whereby parametrising
a moduli space. Three cases need to be addressed separately:
\begin{itemize}
\item $N_f>N_c$: 
The constraints are not modified by quantum correction; 
\item $N_f=N_c$: 
The only classical constraint is modified by quantum effects as
\be 
\label{quantum_correction}
\det(M)-(*B)(*\widetilde{B})=0 \Rightarrow 
\det(M)-(*B)(*\widetilde{B})=\Lambda^{2N_c} \ ,
\ee
where $*$ is the contraction of all flavor indices with the totally
antisymmetric tensor on $N_f$ indices.
\item $N_f<N_c$: Only the $M_i^j$'s exist and they are independent
variables in the moduli space.
However, quantum correction will generate the famous Affleck-Dine-Seiberg
super-potential 
\be
\label{ADS}
W=(N_c-N_f) ({\Lambda^{3N_c-N_f} \over \det(M)})^{1\over N_c-N_f} \ .
\ee 
The reason why the term (\ref{ADS}) can only be
	generated\footnote{$\Lambda$ is the dynamical 
	scale of the  
	asymptotically free (AF) theory. When the energy scale is less
	than $\Lambda$, the gauge coupling becomes strong. 
	So when $\Lambda \rightarrow 0$,
	the gauge theory is weakly coupled at any energy scale
	and there should not be any quantum corrections.}
in the case $N_c>N_f$ is that
when $\Lambda \rightarrow 0$,
(\ref{ADS}) will become singular
if $N_c< N_f$. 
\end{itemize}

Since for $N_c\geq N_f$, there are complicated constraints among 
variables $M_i^j, B, \widetilde{B}$ (i.e., non-trivial moduli space)
which make the problem less tractable, 
we will consider the  simpler case by 
adding some mass terms such that the remaining massless 
flavors are less than $N_c$. For example, we set only
\[
m_j \ne 0 \qquad j=K+1,...,N_f
\]
with $K<N_c$ and add a term into the 
superpotential as 
\be 
\label{electric_deg}
W_{elec}= \sum_{j=K+1}^{N_f}  Q_j m_j \widetilde{Q}^j \ .
\ee 
We shall call (\ref{electric_deg}) a degenerate mass deformation in
contrast to \cite{Feng} where all $m_j$ were non-zero.

After integrating out these massive flavors the theory becomes effectively
$SU(N_c)$ with $K$ flavors, so the exact effective superpotential is 
\be 
\label{ADS_1}
W=(N_c-K) ({\hat{\Lambda}^{3N_c-K} \over \det(M)})^{1\over N_c-K}
\ee 
where $M_i^j, i, j=1,...,K$ is the meson constructed from the
remaining massless flavors and the cut-off scale 
$\hat{\Lambda}$ in IR matches the 
$\Lambda$ in UV by 
\be  
\label{match}
\hat{\Lambda}^{3N_c-K}=\det(m)\Lambda^{3N_c-N_f} \ .
\ee

After refreshing the reader's memory with the above review, 
we can set up our Seiberg dual pair under
the degenerate mass deformation. The electric theory is 
$SU(N_c)$ with $N_f$ flavors and superpotential (\ref{electric_deg})
with $K<N_c$. The corresponding magnetic theory is $SU(N_f-N_c)\equiv
SU(\widetilde{N}_c)$
with $N_f$ flavors, a meson $X_i^j, i,j=1,...,N_f$ and
superpotential 
\be \label{dualdefo}
W_{mag}={1\over \mu}X_i^j q_j  \widetilde{q}^i+ \Tr(Xm)
\ee
where $\mu$ is a scale and $m$ is same mass matrix in (\ref{electric_deg}).
Our aim is to do the matrix model integration for both electric 
and magnetic theories and to show that they reproduce the results
(\ref{ADS_1}) and (\ref{match}).
%
\section{The Matrix Model}
\subsection{The Electric Side}
We first do the matrix model integrations for the electric field
theory.
Unlike the case discussed in \cite{Feng},
here the mass matrix is degenerate and there are some massless fields
left in the IR. To do it correctly, we must take care of these
zero modes as emphasized in \cite{Berenstein:2002sn}. The
way to do this is advocated in \cite{Demasure:2002sc} where we add
a delta-function into the matrix model integration so that we are only
integrating  fields subjects to the constraint given by the moduli
space equation of corresponding field theory. 

More concretely, in our example, we have a $U(N)$
matrix model at large $N$ (we emphasize that at the stage of computing
free-energies, this $N$ is unrelated to the $N_c$ and $N_f$ of the
field theory), and we
should modify the na\"{\i}ve
integration for the partition function
\bean
Z &= & {1\over \vol(U(N))}\int \prod_{j=1}^{N_f} d Q_j dQ^{\dagger}_j
 e^{-{1\over g_s} \sum\limits_{j=K+1}^{N_f} Q_j m_j Q^{\dagger j}}\\
& & =
 {1\over \vol(U(N))}\int \prod_{j=1}^{K} d Q_j dQ^{\dagger}_j
\int \prod_{l=K+1}^{N_f} d Q_l dQ^{\dagger}_l
 e^{-{1\over g_s} \sum\limits_{j=K+1}^{N_f} Q_j m_j Q^{\dagger j}}
\eean
into the form
\be \label{matrix_elec}
Z = {1\over \vol(U(N))}\int \prod_{j=1}^{K} d Q_j dQ^{\dagger}_j
\delta(M_i^j- Q_i Q^{\dagger j})
\int \prod_{l=K+1}^{N_f} d Q_l dQ^{\dagger}_l
 e^{-{1\over g_s} \sum\limits_{j=K+1}^{N_f} Q_j m_j Q^{\dagger j}}
\ee
where we have split the integration into massive and massless
parts and inserted the delta-function constraint in light of the
fact that the meson is composed of the fundamental squarks.

We wish to emphasize that the above treatment should apply for more
general cases such as $N_f>N_c$ without any mass deformations. 
We only need
to include proper delta-function constraints into the matrix model 
integration. These cases without tree-level superpotential which
seemingly defy the rules of the Dijkgraaf-Vafa Programme can thus be
addressed.
The difficult part is that when we include these
delta-functions, it is hard to do the integration in general. 
Developing this technique will be very important to the
matrix model - field theory correspondence.

For the integral (\ref{matrix_elec}) 
there are three contributions. The volume contributes a factor of
$\vol(U(N)) = N^{\frac{N^2}{2}} = 
e^{{N^2\over 2} \log N}=e^{{1\over g_s^2} {S^2\over 2} \log {S\over
g_s}}$.
As we argued in \cite{Feng}, to get the proper dimensions, we need to
replace $g_s=\Lambda^3 e^{3/2}$ and get the contribution to the
effective superpotential as
\be
\label{vol_ele}
\frac{\partial (\frac{S^2}{2} \log \frac{S}{g_s})}{\partial S} =
N_c[S\log {S\over  \Lambda^{3}}-S] .
\ee

The second piece comes from the massive field integration and it is as
in \cite{Feng}:
$$
e^{ N (N_f-K) \log(\pi g_s)- N\log(\det(m))}=
e^{{1\over g_s} [ S (N_f-K) \log(\pi g_s) - S\log(\det(m))]}.
$$ 
Again,
dimensional analysis allows us to replace $\pi g_s$ by $\Lambda$ and
we get the next contribution to the effective superpotential:
\be 
\label{mass_ele}
[ S (N_f-K) \log(\Lambda) - S\log(\det(m))] \ .
\ee

The third piece comes form the integration of the massless modes
subject to the delta-function constraint. The 
contribution is $e^{-K N \log N +(N-K) \log \det(M)}$
\cite{Demasure:2002sc} with the method of Wishart models. Since in
the matrix model, we need to take $N\rightarrow \infty$ and  
$N-K\sim N$ in the second term. When translating into the
field theory, we need to put back the proper dimensionful parametres 
as before to obtain the last contribution:
\beq 
\label{cons_elec}
-K[S\log {S\over  \Lambda^{3}}-S]+ S\log ( {\det(M) \over
\Lambda^{2K}}) \ .
\eeq

Adding the three pieces (\ref{vol_ele})  (\ref{mass_ele}) 
(\ref{cons_elec}) together, we get
\bea
-W_{elec;~eff} & = & N_c[S\log {S\over  \Lambda^{3}}-S]+
[ S (N_f-K) \log(\Lambda) - S\log(\det(m))]  \nonumber \\
& & -K[S\log {S\over  \Lambda^{3}}-S]+ S\log ( {\det(M) \over
\Lambda^{2K}}) \nonumber \\ \label{elec-W}
& = & (N_c-K)[ S \log { S \over ( {\det(m) \Lambda^{3N_c-N_f} \over
\det(M) })^{1\over N_c-K}} -S] \label{W_elec_S} \ .
\eea
Minimizing (\ref{W_elec_S}) with respect to $S$ we get 
\be 
\label{S_elec}
S=  ( {\det(m) \Lambda^{3N_c-N_f} \over \det(M) })^{1\over N_c-K}
\ee 
so the exact superpotential, upon integrating out $S$ is
\be 
\label{elec_no_S}
W_{elec;~eff}= (N_c-K) ( {\det(m) \Lambda^{3N_c-N_f} \over
\det(M) })^{1\over N_c-K} \ ,
\ee
which is exactly the result in the field theory (\ref{ADS_1}) and
(\ref{match}).

As we emphasized above, this calculation is suitable only for 
the case of $K<N_c$  because only in this case, the 
independent variables are just $M_i^j$ and there are no baryonic
fields; this is reflected in our matrix calculation since we would
otherwise need to insert extra delta-function constraints to capture
the moduli space. Indeed
this restriction also gives a hint of how the matrix model actually
sees different behavior
in the field theory for the cases $N_c> N_f$ and $N_c\leq N_f$ because
of the necessity of putting in different constraints.
%
\subsection{The Magnetic Side}
Now let us move to the dual magnetic side.  The tree-level
superpotential is given
in (\ref{dualdefo}). Again, since 
the mass matrix $m$ is degenerate,  we need to modify the
matrix model integration by including the proper delta-function as
in the previous subsection.
In particular, we have
\be 
\label{matrix_mag}
Z = {1\over \vol(U(N))}\int dX \prod_j dq_j dq^{\dagger}_j
[\prod_{i,j=1}^K\delta(X_i^j-M_i^j)] \exp(
{-1\over g_s}[  \Tr(m X)+ \sum_{i,j=1}^{N_f} {1\over \mu}
 X_i^j q_j  q^{\dagger i}]) \ .
\ee
After finishing the integration of $X$ as was done in \cite{Feng}, 
(\ref{matrix_mag}) becomes
\be 
\label{red_mag}
Z = {1\over \vol(U(N))} \int \prod_{j=1}^K  dq_j dq^{\dagger}_j
e^{-{1 \over g_s \mu} \sum\limits_{i,j=1}^K M_i^j q_j  q^{\dagger i}}
\int \prod_{l=K+1}^{N_f}  dq_l dq^{\dagger}_l 
\delta(\mu m_l^p+ q_l  q^{\dagger p}) \ .
\ee

Once again, the whole integration (\ref{red_mag}) is reduced to 
three contributions. The first one comes from the
volume and as in \eref{vol_ele} gives the contribution to the
superpotential as
\be 
\label{vol_mag}
\widetilde{N}_c[S\log {S\over  \widetilde{\Lambda}^{3}}-S] \ ,
\ee
where we use $\widetilde{N}_c$, $\widetilde{\Lambda}$
to indicate that it is in the dual magnetic theory. 

The second piece is simply the Gaussian integration for massive fields
because here $M_i^j$ are just the mass parameters and upon comparison
with (\ref{mass_ele}) we obtain the contribution
\be
\label{mass_mag}
[ S K \log( \widetilde{\Lambda}) - S\log(\det({M_i^j \over \mu}))] \
.
\ee

The third piece is the same constrained integration as given by
\cite{Demasure:2002sc} 
and the contribution is (comparing with (\ref{cons_elec}))
\be \label{cons_mag}
-(N_f-K)[S\log {S\over   \widetilde{\Lambda}^{3}}-S]
+ S\log ( {\det(-\mu m) \over \widetilde{ \Lambda}^{2(N_f-K)}})
\ .
\ee

It is interesting to notice that the mass integration (\ref{mass_mag}) 
and constraint integration (\ref{cons_mag}) in the magnetic field theory 
are exactly the
opposite of the corresponding electric field theory (\eref{mass_ele}
for the mass and \eref{cons_elec} for the constraint integrals). 
This of course is no coincidence and is in fact a result of Seiberg
duality. 

Putting the three pieces together we get
\bean
-W & = & \widetilde{N}_c[S\log {S\over  \widetilde{\Lambda}^{3}}-S]
+[ S K \log( \widetilde{\Lambda}) - S\log(\det({M_i^j \over \mu}))] \\
& & \qquad -(N_f-K)[S\log {S\over   \widetilde{\Lambda}^{3}}-S]
+ S\log ( {\det(-\mu m) \over \widetilde{\Lambda}^{2(N_f-K)}}) \\
& = & (\widetilde{N}_c-(N_f-K))[S\log {S\over  \widetilde{\Lambda}^{3}}-S]
+ S\log { (-)^{N_f-K} \mu^{N_f} \det(m) \over \det(M) 
\widetilde{\Lambda}^{2N_f-3K}} \\
& = & (\widetilde{N}_c-(N_f-K))[S \log{ S\over  \widetilde{\Lambda}^{3}
({ (-)^{N_f-K} \mu^{N_f} \det(m) \over \det(M) 
\widetilde{\Lambda}^{2N_f-3K}})^{-1\over\widetilde{N}_c-(N_f-K)}}-S] \ . 
\eean 

Minimizing the above superpotential with respect to $S$ we obtain
\be
\label{S_mag}
S=\widetilde{\Lambda}^{3}
({ (-)^{N_f-K} \mu^{N_f} \det(m) \over \det(M) 
\widetilde{\Lambda}^{2N_f-3K}})^{-1\over\widetilde{N}_c-(N_f-K)}
\ ;
\ee
after some algebra it can be shown that $S_{mag}=-S_{ele}$.
The exact superpotential, upon back substitution becomes
\bea
W_{mag;~eff} &  = & (\widetilde{N}_c-(N_f-K))\widetilde{\Lambda}^{3}
({ (-)^{N_f-K} \mu^{N_f} \det(m) \over \det(X) 
\widetilde{\Lambda}^{2N-f-3K}})^{-1\over\widetilde{N}_c-(N_f-K)}
\nonumber \\
& = & (\widetilde{N}_c-(N_f-K))({ (-)^{N_f-K} \mu^{N_f} \det(m) \over \det(X) 
\widetilde{\Lambda}^{3\widetilde{N}_c-N_f}})^{-1\over\widetilde{N}_c-(N_f-K)}
\ .
\label{mag_no_S}
\eea

Now using the relationships \cite{Argyres} 
\be \label{relation}
 \Lambda^{3N_c-N_f} \widetilde{\Lambda}^{3\widetilde{N}_c-N_f}
=(-)^{N_f-N_c} \mu^{N_f},~~~~\widetilde{N}_c-(N_f-K)=-(N_c-K)
\ee
for the dual cut-off scales, we can recast (\ref{mag_no_S}) into
\be  \label{mag_change}
W_{mag;~eff}  =  (N_c-K) ( { \det(m) \Lambda^{3N_c-N_f} \over
\det(M)})^{1\over N_c-K} \ ;
\ee
note that the minus signs have been properly canceled. 

We recognize (\ref{mag_change}) as precisely (\ref{elec_no_S});
therefore the matrix model computation has again successfully
reproduced Seiberg duality in this generalized case from the one in 
\cite{Feng}. 
%
\section{Discussions and Prospects}
In this paper, we have generalized  
the result in \cite{Feng}, from 
non-degenerate to degenerated mass matrix and
have shown that in the context of the Dijkgraaf-Vafa Programme, the 
matrix model continues to perfectly reproduce the predictions of 
Seiberg duality. 

The techniques arising from this illustrative example extend beyond
the present framework. In fact they allow us to
propose a general method of attack on the
matrix model integration when the corresponding field theory has a 
classical moduli space, by generalizing the ideas presented in 
\cite{Berenstein:2002sn,Demasure:2002sc}. 

In particular, we need to add into the partition function integral,
proper delta-function constraints in accordance with the explicit
relations in the field theory moduli space. In other words, one cannot
na\"{\i}vely integrate over the space of all matrices but only
subspaces relevant to the field theory. It is worth to emphasize
that these constraints we add are {\sl classical} relationships.
The matrix model will supply the quantum correction to the
moduli space. This can be seen by setting $K=N_c$ in equation
(\ref{elec-W}), so the equation of motion of $S$ gives
$\det(M)-\det(m)\Lambda^{3N_c-N_f}=0$. Because for different
number of flavors we will have different delta-functions,
this prescription
solves the puzzle why the matrix model would
know the different dynamical behavior of the corresponding
field theory. 

Moreover, field theories without tree-level
superpotential which {\it ab initio} seemingly elude the
Dijkgraaf-Vafa procedure, 
can be thus addressed. Indeed we merely have to add
appropriate delta-function constraints (and go to the dual
electric/magnetic theory if necessary) to perform the matrix integral.

However, as remarked in \cite{Feng}, we are still far
from completely showing Seiberg duality in the matrix model, 
even for the
standard example of no mass deformations at all.
The difficulty is that we need to find proper
delta-function constraints reflecting the baryonic and mesonic
branches of the moduli space, and more importantly, to do the
matrix integration in the presence of these constraints.
This is a very involved task and beckons for future work.

Many immediate checks are also conveniently at hand.
The generalized Seiberg dualities, such as the host of examples in 
toric dualities and quiver dualities addressed in
\cite{Feng:2000mi,Cachazo:2001sg} and \cite{Berenstein:2002fi} present
as readily available case-studies.
It is also
interesting to generalize our treatment from $U(N)$
to $SO/Sp$ gauge groups.

The works in \cite{Feng} and herein are a nontrivial
check of Seiberg duality in matrix models. However, 
we would like to ask
a more profound question: could we derive Seiberg duality from the
matrix model? In other words, we start with a known electric
field theory and translate it into the proper matrix model. Then could
we find a transformation in the matrix integration to
change this electric matrix model into another
equivalent magnetic one, from which we can read out the superpotential
of the magnetic field theory directly?
By this way, we would have {\sl derived} Seiberg duality from the matrix 
model and be granted the remarkable ability to see an ${\cal N}=1$
duality purely from a bosonic matrix integration.

There are some hints for this
interesting issue in our calculations. Comparing 
(\ref{matrix_elec}) and (\ref{red_mag}), we see the constrained
integration in one model becomes direct integration in another
and vise versa. It is reminiscent of some kind of field theory
transformation with source such as Legendre transformations.
Does this hold in general? What is this transformation in the matrix
model which we seek that would derive Seiberg Duality?
\section*{Acknowledgements}
We are grateful to Vijay Balasubramanian,
David Berenstein, Freddy Cachazo, Joshua Erlich,  
Min-xin Huang, Vishnu Jejjala, Asad Naqvi and Nathan Seiberg for
enlightening conversations. We are also indebted to the gracious
patronage of the Institute for Advanced Study as well as the Dept.~of
Physics at the University of Pennsylvania.
This research is supported in part 
under the NSF grant PHY-0070928 (BF) and
the DOE grant DE-FG02-95ER40893 (YHH).
\bibliographystyle{JHEP}

\end{document}